\documentclass[aps, prb, twocolumn, superscriptaddress, longbibliography]
    {revtex4-1}

\usepackage{graphicx}
\usepackage{epstopdf}
\usepackage{dcolumn}
\usepackage{bm}
\usepackage{color}
\usepackage{natbib}

\begin{document}


\title{Band structure of a HgTe-based three-dimensional topological insulator}

\author{J. Gospodari\v{c}}
\author{V. Dziom}
\author{A. Shuvaev}
\affiliation{Institute of Solid State Physics, Vienna University of
Technology, 1040 Vienna, Austria}
\author{A. A. Dobretsova}
\author{N.~N.~Mikhailov}
\author{Z. D. Kvon}
\affiliation{Rzhanov Institute of Semiconductor Physics and
Novosibirsk State University, Novosibirsk 630090, Russia}
\author{E. G. Novik}
\affiliation{Dresden High Magnetic Field Laboratory (HLD-EMFL), Helmholtz-Zentrum Dresden-Rossendorf, 01328 Dresden, Germany}
\affiliation{Institute of Theoretical Physics,
Technische Universit\"{a}t Dresden, 01062 Dresden, Germany}
\author{A. Pimenov}
\affiliation{Institute of Solid State Physics, Vienna University of
Technology, 1040 Vienna, Austria}

\begin{abstract}
From the analysis of the cyclotron resonance, we experimentally obtain the
band structure of the three-dimensional topological insulator based on
a HgTe thin film. Top gating was used to shift the Fermi level in the film,
allowing us to detect separate resonance modes corresponding to the surface
states at two opposite film interfaces, the bulk conduction band, and the
valence band.
The experimental band structure agrees reasonably well with the predictions
of the $\mathbf{k\cdot p}$ model. Due to the strong hybridization of the surface
and bulk bands, the dispersion of the surface states is close to parabolic
in the broad range of the electron energies.

\end{abstract}

\date{\today}


\maketitle

\section{Introduction}\label{Intro}

The electronic band structure provides an important fingerprint of a material
in the reciprocal space. In case the surface of the sample is accessible
experimentally, the standard technique of angle-resolved photoemission spectroscopy~\cite{damascelli_rmp_2003} is an established way to obtain the
necessary information. However, in several cases, especially in two-dimensional
heterostructures, several buffer or capping layers prevent collecting the data
from the photoemitted electrons.
As possible alternative methods, the analysis of the cyclotron
mass~\cite{novoselov_nature_2005, zhang_nat_2005, zhang_nphys_2011,
minkov_prb_2014} or density of states via capacitance
experiments~\cite{kozlov_prl_2016, kozlov_jetpl_2016} have been suggested
to recover the band dispersion especially of two-dimensional materials.
In magneto-optical experiments~\cite{hancock_prl_2011, orlita_np_2014,
zoth_prb_2014, dantscher_prb_2015, akrap_prl_2016}, the relevant information
is obtained comparing the theoretical predictions~\cite{chu_book, yu_book}
of the band structure with experimental data.

Within another approach, the band structure may be obtained from the analysis
of the cyclotron resonance frequencies~\cite{minkov_prb_2014, shuvaev_prb_2017}
that is especially useful for two-dimensional materials. Indeed, in two
dimensions and in the quasiclassical approximation, the cyclotron frequency
$\Omega_c$ may be written in terms of the cyclotron effective mass $m_c$
as~\cite{ashcroft_book}
\begin{equation}
m_c \equiv \frac{eB}{\Omega_c} =
	\frac{\hbar^2}{2 \pi }\frac{\partial A}{\partial E} \Big|_{E=E_F} \ .
\label{eqCR}
\end{equation}
Here, $B$ is the external magnetic field, $A$ is the area in the reciprocal
space enclosed by the contour of the constant energy $E$, and $E_F$ is the
Fermi energy.

An important point is that the cyclotron
frequency in Eq.\,(\ref{eqCR}) linearly depends on external magnetic field independently
of the form of dispersion relations because the Fermi area is
fieldindependent in the quasiclassical approximation. This approximation
is the main assumption in the present experiments~\cite{minkov_prb_2014,
shuvaev_prb_2017}, i.e., transitions between several Landau levels should
take place simultaneously. This condition is certainly realized at lower
magnetic fields utilized in the present experiment.

In the general case, the relation between the area and the band structure may be complicated.
In such cases, an additional input from the theory is indispensable. As discussed in
Sec. \ref{secmod}, especially for surface states and for bulk conduction band,
the isotropic approximation can be applied leading to a simple relation between
the Fermi-vector $k_F$ and the Fermi area:
$A=\pi k_F^2$. In this case, Eq.\,(\ref{eqCR}) can be rewritten as
\begin{equation}
\frac{\partial E}{\partial k} \Big|_{E=E_F} = \frac{\hbar^2 k_F}{m_c}\ ,
\label{eqCR2}
\end{equation}
and, thus, can be directly integrated to obtain the experimental band structure
$E(k)$. For the holelike states, however,
the isotropic approximation breaks down, and additional information from the
theory is necessary to obtain the band structure. A possible
approach, in this case, is presented in Sec.~\ref{secmod}.

In this paper, we apply the procedure sketched above to a three-dimensional
topological insulator (3D TI) HgTe and compare the results with the
predictions of the $\mathbf{k \cdot p}$ model.

\begin{figure}[th!]
	\includegraphics[width=1\linewidth]{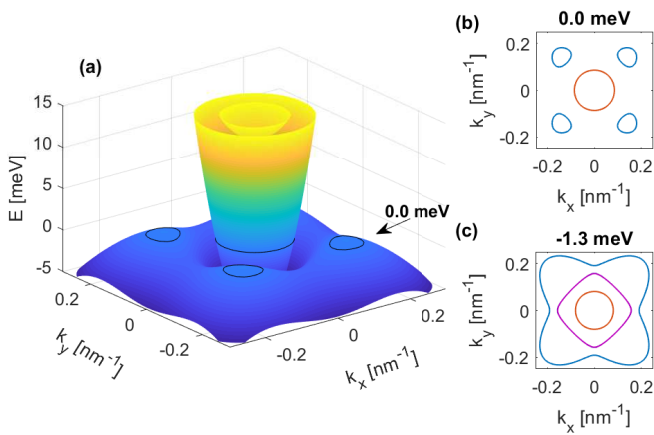}
	\caption{\textit{Theoretical band structure of a 3D topological insulator
			HgTe.} (a) The first valence band, the surface band, and the first conduction band of the 80-nm-thick HgTe
		layer at the charge neutrality point $n_{tot}=0$ where the concentrations of electrons
		and holes equal  $\pm1.2 \cdot 10^{11}\text{~cm}^{-2}$.
		(b) Cross section of (a) at $E_f=0$~meV.
		Blue: holelike Fermi surface ($\partial A /\partial E <0$) from the islands in the $(\pm1,\pm1)$ directions. Orange:
		electronlike Fermi surface ($\partial A /\partial E > 0$) from the surface states. (c) Fermi surface of the hole-doped sample
		where four islands are connected and lead to a different cyclotron picture:
		blue: holelike, violet: electronlike, orange: electronlike.}
	\label{figth}
\end{figure}

The unstrained single-crystalline mercury telluride (HgTe) is a gapless
semimetal with the conduction and valence bands formed by $\Gamma_8$
bands~\cite{chu_book}. If grown in the form of a thin film on a CdTe layer,
HgTe is subject to a tensile strain due to lattice mismatch between HgTe and
CdTe~\cite{brune_prl_2011}. As a consequence, the originally degenerate
light and heavy $\Gamma_8$ hole bands split at the $\Gamma$ point,
thus, forming a bulk insulator with
a gap around $\sim 20$\,meV~\cite{brune_prl_2011,kozlov_prl_2014} for an
$80$-nm HgTe film. Due to a band inversion between HgTe and CdTe the
topologically protected surface states arise in about 10-20-nm thick layer
close to the boundary. HgTe films, thus, form a strong 3D TI~\cite{fu_prb_2007}. According to band-structure
calculations, the conduction band of an $80$-nm HgTe 3D TI is nonparabolical,
and it is quantized due to the confinement. The valence band of the HgTe
film reveals a deep minimum at the $\Gamma$ point with four shallow side
maxima along the $(\pm1, \pm1)$ directions (see Fig.\,\ref{figth}).
The minimum is due to the mixing between light and heavy-hole states in the inverted band structure of
HgTe~\cite{ortner_prb_2002, dziom_ncomm_2017}.

\section{Experiment}
\label{secexp}
\subsection{Technique}

The magneto-optical experiments were carried out on a strained
$80$-nm-thick HgTe film grown by molecular beam epitaxy on a (100)-oriented
{GaAs} substrate\cite{kvon_ltp_2009, kozlov_prl_2014} with a lateral size of
$5\times5$~mm. The layer was sandwiched between the cap and the buffer layers of
Cd$_{0.7}$Hg$_{0.3}$Te to obtain high electron mobility in the sample.
The analysis of the cyclotron resonance corresponding to the upper surface
state revealed the mobility to be up to
$\mu=e \tau/m_c=5\cdot 10^5$\,cm$^{2}$/Vs. Here, $e$, $\tau$, and $m_c$ are the
electron charge, the scattering time, and the cyclotron mass, respectively.
Between the layered structure and the GaAs substrate a $5$-$\mu$m- thick
CdTe buffer layer was placed.
To produce the semitransparent gate electrode, the film was covered on top by
a multilayer insulator of SiO$_{2}/$Si$_3$N$_4$ and a semitransparent
metallic $10.5$-nm-thick Ti-Au layer. The top-gate electrode allowed the
variation of the Fermi energy to probe the surface states and the valence
and conduction bands~\cite{shuvaev_apl_2013, brune_prx_2014, kozlov_prl_2016}. The shape of
the gate electrode allowed a fully covered center of the sample for terahertz
transmission measurements and four contacts at the corners of the sample,
allowing to acquire simultaneous information about the electrical
conductivity in the system.

Due to the fact that the gate only partially covers the sample,
 we were not able to fully rely on the magnetotransport measurements.
The ungated regions can significantly falsify the transport response. However,
at zero gate voltage, this effect is minimized, thus, allowing us
to gather some additional information about the carriers in
the system as shown below.

The cyclotron resonance was investigated in a Mach-Zehnder
interferometer~\cite{volkov_infrared_1985} arrangement, which allowed to
measure the amplitude and the phase shift of the transmitted electromagnetic
radiation with controlled polarization of light~\cite{shuvaev_sst_2012,
shuvaev_apl_2013}. Continuous monochromatic radiation was produced by
backward-wave oscillators operating in the submillimeter regime
($100$\,GHz\,--\,$1000$\,GHz). The transmitted radiation
was detected by a silicon $4.2$-K bolometer with a high-frequency
cutoff filter of $3$~THz. The data were obtained at several fixed
frequencies in sweeping magnetic fields. Additional information about the
charge carriers in the system was also obtained from the frequency-dependent
spectra in a zero magnetic field. To unambiguously separate the
resonances from the electronlike and holelike carriers, several experiments
were conducted with circularlypolarized radiation. The experiments were
carried out at $1.8$ K in a split-coil superconducting magnet that provided a
magnetic field up to $\pm 7$\,T in the Faraday geometry;
i.e., a magnetic field was applied parallel to the $\mathbf{k}$ vector of
the terahertz radiation.

\subsection{Spectra modeling}
\label{secspec}

In order to obtain the parameters of the charge carriers, such as the
two-dimensional (2D) density $n$, the effective cyclotron mass $m_c$ and the
intrinsic scattering time $\tau$, the acquired experimental data were fitted
using the Drude model for dynamical conductivity in the quasi-classical
approximation~\cite{palik_rpp_1970}. We utilize the geometry with circularly
polarized light and the conductivity written as
\begin{equation}
\sigma_{+} = \frac{\sigma_0}{1- i\tau (\omega + \Omega_c)} \ .
\label{eq_drude}
\end{equation}
Here $\sigma_0=ne^2\tau/m_c$ is the two-dimensional DC conductivity.
 The
conductivity of a system with multiple carriers is the sum of the individual
conductivities. Neglecting the influence of the substrate, the simplified
expression for the transmission of circularly polarized radiation through
a film (assumed to be thin compared to the radiation wavelength) can be
written as:
\begin{equation}
t_+ =1- \frac{i}{\tau_{\mathrm{SR}}}\frac{1}{(\omega+i \Gamma)-\Omega_c} \ .
\label{eqtr}
\end{equation}
Here the "total" scattering rate
$\Gamma = {1}/{\tau}+ {1}/{\tau_{\mathrm{SR}}}$ describes the effective
width of the cyclotron resonance observed in the transmission signal,
$1/\tau$ is the transport scattering rate,
$1/\tau_{\mathrm{SR}} = ne^2 Z_0/2m_c$ is the superradiant
damping~\cite{gospodaric_prb_2019}, and $Z_0$ is the impedance of the free
space.

To take into account the reflections inside the substrate, we employed a more
accurate model for the analysis of the experimental results. The procedure
utilizes similar algebra as described
previously~\cite{shuvaev_prl_2011, dziom_ncomm_2017, candussio2019cyclotron}.
The theoretical transmission of the circular polarization
$t_+ = t_{xx} +i t_{xy} $ was calculated by changing into the basis with
parallel $t_{xx}$ and crossed $t_{xy}$ transmission coefficients.

\section{Theoretical model}
\label{secmod}

To acquire a more detailed insight into the band structure of the
strained HgTe layer, theoretical calculations have been performed using a
multiband {\bf k}$\cdot${\bf p} model~\cite{novik_prb_2005} which takes into
account the strong coupling between the lowest conduction and the highest
valence bands. The {\bf k}$\cdot${\bf p} model considers eight bands: two $\Gamma_6$,
two $\Gamma_7$ and four $\Gamma_8$ subbands. Yet, considering the energy region of
our interest, the contribution of the
$\Gamma_7$ subband is below $1\%$. The calculations were performed for a fully strained
HgTe film with Cd$_{0.7}$Hg$_{0.3}$Te barriers which are grown on a CdTe substrate.
The strain due to the lattice mismatch between HgTe and CdTe of about 0.3$\%$
leads to an opening of a direct gap of $\approx 22$\,meV (the indirect gap is
about $10-15$\,meV) between the heavy-hole and light-hole bands in the HgTe
layer~\cite{brune_prl_2011}. We take the strain effects into consideration by
applying a formalism introduced by Bir and Pikus~\cite{bir_pikus_1974}.
According to the previous studies of similar
structures~\cite{brune_prl_2011, hancock_prl_2011, wu_eurlett_2014}, the
crossing point of the surface states is located below the bulk band gap.
Accordingly, a full-band envelope function
approach~\cite{andlauer_prb_2009} is used to perform the self-consistent
calculations of the Hartree potential. This procedure avoids the separation
of the occupied electron and hole states which is complicated for structures
where both are occupied simultaneously.
Our calculations include the structure inversion asymmetry and, therefore,
effectively reproduce the experimental effect of the applied gate. The spatial
distribution of charge can be calculated while the total charge density is
being varied. The Hartree potential determined by this spatial distribution
of charge (see Eq.\,(2) in Ref.\,[\onlinecite{andlauer_prb_2009}])
splits
the bulk and surface states and leads to their realignment, resulting in
significant band structure modifications
(see Figs.\,\ref{fig_disp} and \ref{fig_hartree} and the discussion  below).

\begin{figure}[tbp]
\includegraphics[width=0.99\linewidth]{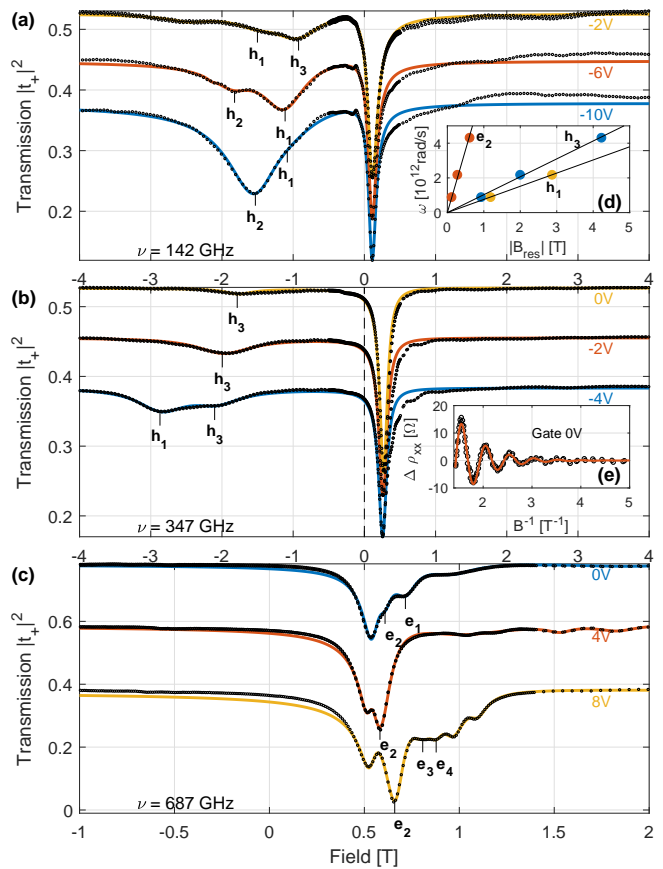}
\caption{\textit{Cyclotron resonance with circularly polarized light.}
(a)-(c) The intensity of the transmitted radiation $|t_+|^2$ as a function of an
external magnetic field for fixed frequencies as indicated. Resonance
features for positive and negative fields correspond to electrons and holes,
respectively. Points: experiment, solid lines: theoretical model based on
Drude conductivity, Eq.\,(\ref{eq_drude}). The absolute scales refer to the
lowest curves, the others are shifted for clarity. The inset (d) shows the field
dependence of the cyclotron resonance demonstrating linear behavior within
the quasiclassical approximation according to Eqs.\,(\ref{eqCR}) and (\ref{eqCR2}).
(e) The oscillating part of the
longitudinal resistivity at zero gate voltage. The experimental data
(black circles) were fitted by a single-carrier Lifshits-Kosevich model (orange line).}
\label{fig_spec}
\end{figure}

\textcolor{black}{
There is an ongoing debate in the literature about the influence of the interface
inversion asymmetry (IIA)~\cite{tarasenko_prb_2015} on the band structure of HgTe structures.
Several studies argue a sufficient effect of IIA in HgTe quantum
wells grown on (013) substrates~\cite{minkov_prb_2016, Minkov2017, bovkun_jpcm_2019}.
Moreover, theoretical calculations in Ref.~[\onlinecite{tarasenko_prb_2015}] predict
a gap of about 15 meV caused predominantly by IIA in (001) HgTe quantum wells of
critical thicknesses. However, experimental data~\cite{buttner_nphys_2011, shuvaev_prb_2017} do not confirm
these predictions. Considering
the complexity of the band
structure of 80 nm HgTe layers and the missing experimental evidence of the influence
of the IIA, we do not include these terms in our calculations.
}

In an attempt to test further anisotropy terms in the Hamiltonian, we tried to include the bulk inversion asymmetry (BIA)~\cite{dresselhaus_pr_1955} term in calculation of the band structure at the charge neutrality point ($n_{tot}=0$). As demonstrated in Fig.\,S9 of the Supplemental Material in Ref.~[\onlinecite{supp}], the inclusion of this term strongly splits the valence bands and reduces the value of the gap. As the latter even worsens the agreement between theory and experiment (see Fig.\,\ref{fig_disp} below), the BIA term was not used in the calculations of the band structure.

The plots of the surface band, the first valence band, and the first conduction band
calculated using the
{\bf k}$\cdot${\bf p} model are shown in Fig.\,\ref{figth} for the case of
the charge neutrality point: The densities of the holes and electrons are equal,
and the total charge density equals $n_{tot}=0$. Here, all three
bands are spin degenerate. In this case, the Fermi
energy crosses the surface, and the valence bands, thus, forming four islands
in the valence band as shown in Fig.~\ref{figth}(b).

In the cyclotron signal, we expect an electronlike resonance due to the
surface states and a holelike signal from the valence islands.
After lowering the Fermi energy, the four regions of the valence band connect
forming a ring structure as shown in Fig.~\ref{figth}(c). In this case, the
fourfold "valley" degeneracy is lifted, and each curve of the valence-band
ring corresponds to a separate cyclotron resonance: a holelike signal from
the outer curve (blue) and an electronlike signal from the inner curve
(violet). The latter effect is due to a different sign of
$\partial A/\partial E$ in Eq.\,(\ref{eqCR})
; positive curvature: electrons; negative curvature: holes.
We note that, even in this case,
a separate surface resonance is expected that remains electronlike.
The island-ring transition is present even if we take into account that
the band structure deforms with varying density and that all bands are
spinpolarized due to broken symmetry.
After the Fermi surface is determined
from the band-structure calculations, the theoretical cyclotron mass can be
calculated using the definition in Eq.\,(\ref{eqCR}).

To obtain the density dependence of the cyclotron mass within the present theory  the effect of the applied gate  was modeled by varying the total charge density
in the system from $6 \cdot 10^{11}\text{~cm}^{-2}$ (holes) to
$-6 \cdot 10^{11}\text{~cm}^{-2}$ (electrons) with the Fermi level reaching
the valence and conduction subbands, respectively. For each value of the $n_{tot}$, the cyclotron mass was calculated using Eq.\,(\ref{eqCR}) as a function of density within the corresponding bands.

Finally, the theoretical band structure confirms
the rotational symmetry of both surface states and bulk conduction subbands,
thus justifying the use of Eq.\,(\ref{eqCR2}) to connect the Fermi vector and
the cyclotron mass.
On the other hand, the hole islands do not show this isotropic behavior.
Nevertheless, at lower hole concentration the islands can be approximated as
circles [see Fig.~\ref{figth}(b)] with an effective radius $k_{eff}$ shifted by
$k_0\approx (\pm 0.15, \pm 0.15)\,\text{~nm}^{-1}$ from the $\Gamma$ point.
In this case,
$k=k_{eff}$ in Eq.~(\ref{eqCR2}), where $k_{eff}$ is related to the Fermi-surface area
of each of the four islands as $A=\pi k_{eff}^2$.
Of course, the exact relation between $A$ and $k_{eff}$ can be
calculated from the theory. We believe, however, that a reasonable picture
of the band structure can be obtained within an isotropic approximation as well.
A direct comparison between theory and experiment can be performed using
an approximation-independent plot of cyclotron masses vs. density
(see Fig.\,\ref{fig_mass} below). This presentation is not sensitive to approximations performed in Eq.\,(\ref{eqCR2}).

\section{Results and discussion}
\label{secres}

\subsection{Cyclotron resonance}

Figure~\ref{fig_spec} shows typical field-dependent transmission in the
geometry with circularly polarized radiation. The advantage of this geometry
is the clear separation of the electron ($\mathbf{e}$) and hole ($\mathbf{h}$)
resonances as they are observed for positive and negative external magnetic
fields, in agreement with Eq.\,(\ref{eq_drude}). The inset in
Fig.\,\ref{fig_spec} demonstrates linear field dependence of the cyclotron
resonance frequency ($\omega=2\pi \nu$), thus, verifying the application of the
quasiclassical approximation in Eqs.\,(\ref{eqCR}) and (\ref{eqCR2}). The data at
low frequency (142\,GHz) shown in Fig.\,\ref{fig_spec}(a) are most sensitive
to the overall behavior of the charge carriers as, here, electrons and holes
may be easily observed simultaneously, and they, indeed, can be well separated
for the $U_g=-10$-V curve. For large negative voltages, the Fermi energy is
situated in the valence band. The cyclotron signal from the holelike carriers
can be observed in the gate voltage range from $-10$\,V to $0$\,V. This correlates
with the position of the charge neutrality point that has been estimated from
the resistivity measurements: The longitudinal resistivity $\rho_{xx}$
showed a maximum at around $-3$\,V. With increasing gate voltage, the single
resonance of the electrons reveals a distinct structure that is most clearly
seen in the data at 687\,GHz, Fig.\,\ref{fig_spec}(c).

The transmission curves can be fitted well using the procedure presented in
Sec.\ref{secspec} (solid lines in Fig.\,\ref{fig_spec}). From the analysis of
the resonances in the transmission, we obtain the 2D charge density, effective
cyclotron mass, and the scattering time for each separate carrier type.
A gradual increase in density with
increasing gate voltage is expected for electrons.
Similarly, the density of the
holelike carriers must be a decreasing function of the gate voltage.
Therefore, in the analysis of the band structure, only the resonances caused
by carriers with monotonous gate-voltage dependence of the charge density
were taken into account. For completeness, the electrodynamic parameters of
the remaining resonances are given in the Supplemental Material in Ref.~[\onlinecite{supp}]. We believe
that the majority of the additional peaks represents direct transitions
between Landau levels and, thus, cannot be described via the quasiclassical
approximation using Eqs.\,(\ref{eqCR}) and (\ref{eqCR2}). For example, carriers
$\mathbf{h_1}$ and $\mathbf{h_3}$ in Fig.\,\ref{fig_spec} showed a
non-monotonous gate-voltage dependence of density and were, therefore, not
considered in the band-structure analysis. Nevertheless, we were able to
recognize them at multiple frequencies, showing a characteristic behavior
of charged carriers in our model. Currently, the gate-voltage dependence of
the intensity of these modes cannot be used to extract their density since
their behavior goes beyond the quasiclassical approach.

\begin{figure}[tbp]
\includegraphics[width=1\linewidth]{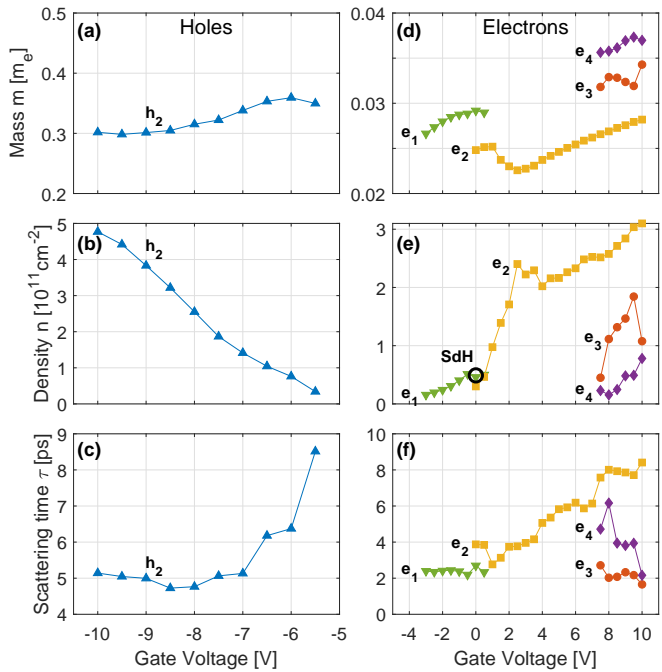}
\caption{\textit{Electrodynamic parameters of the cyclotron resonances in
HgTe.} (a)-(c) -- Holelike carriers, (d)-(f) -- electronlike carriers. Only
the most relevant resonances which may be explained via quasiclassical
picture are shown. Colored symbols are experimental data from the fits of the spectra
in Fig.\,\ref{fig_spec}. The black circle corresponds to the density  $\frac{n_{SdH}}{2}$
resulting from the Shubnikov-de Haas analysis from Fig.\,\ref{fig_spec}(e).
 The lines are guides to the eye.}
\label{fig_par}
\end{figure}

\begin{figure*}[tbp]
\includegraphics[width=0.75\linewidth]{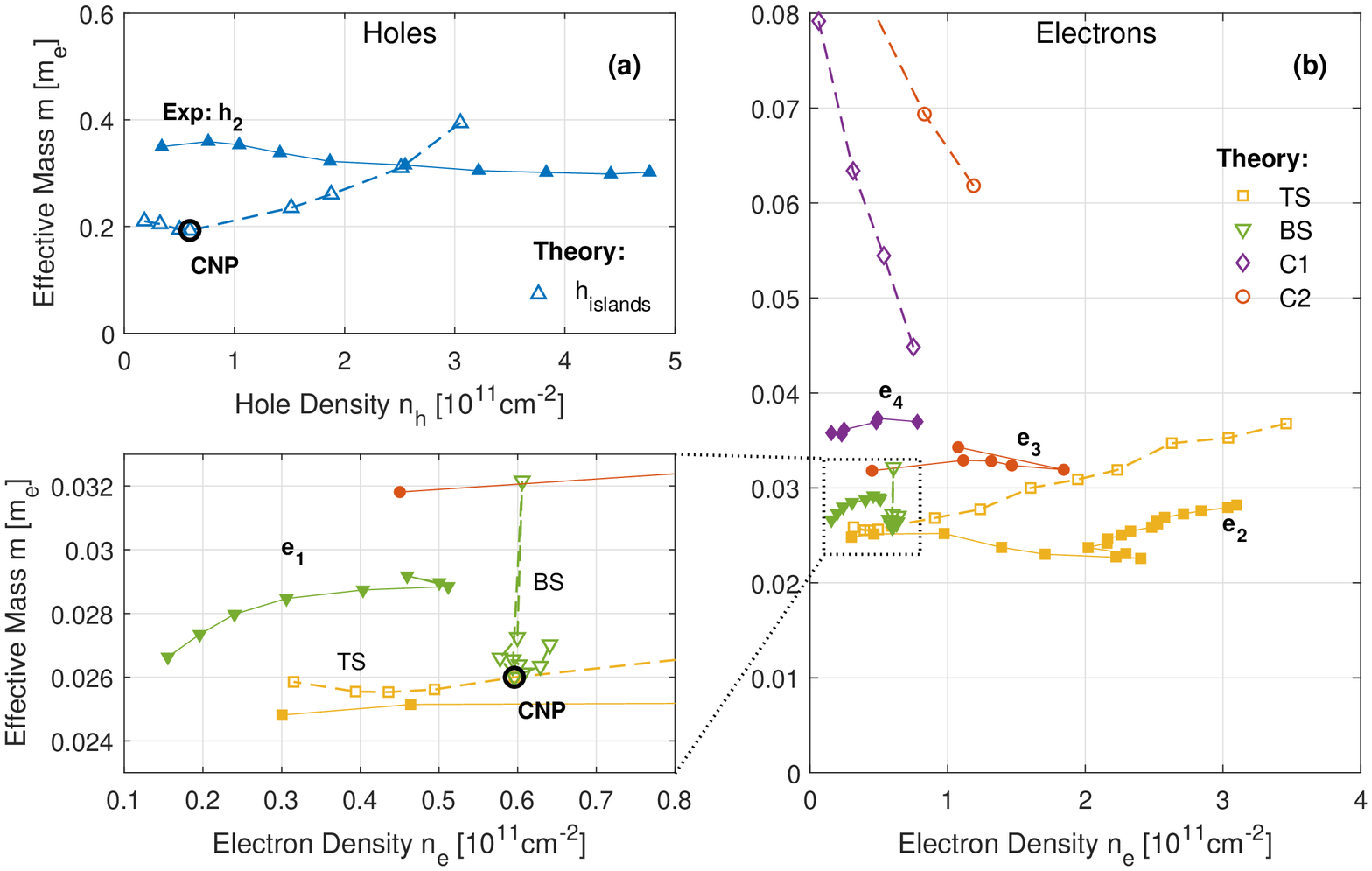}
\caption{\textit{Comparison of the cyclotron masses in strained HgTe with
$\mathbf{k \cdot p}$ model calculations.} Compared to Fig.\,\ref{fig_par},
the cyclotron masses are plotted as a function of density. This presentation
allows the comparison with the theoretical model without integrating
Eq.\,(\ref{eqCR2}). (a) Holelike carriers. (b) Electronlike carriers.
Full symbols: experimental values; empty symbols: theory; BS: bottom
surface states; TS: top surface states; C1, C2: spin-polarized bulk
conduction bands; CNP: charge neutrality point.}
\label{fig_mass}
\end{figure*}

Figure \ref{fig_par} shows the parameters of the cyclotron resonances that
will be used to obtain the band structure of the 3D TI.
The charge density decreases with the gate voltage for holes and increases
for electrons. Both agree with the sign of the charge carriers obtained
directly from the spectra in Fig.\,\ref{fig_spec}.

Additional information about the carriers in the sample was gathered by
four-point longitudinal resistivity measurements at zero gate voltage,
which displayed strong SdH oscillations as plotted against the reciprocal magnetic field in Fig.\,\ref{fig_spec}(e).
The Lifshits-Kosevich
formula\,\cite{Lifshits1958,Minkov2017, Dobretsova2019} can be used to extract the carrier properties from the
oscillation period. A model with a single carrier type fits the experimental data
reasonably well [see the orange curve in Fig.\,\ref{fig_spec}(e)]. The oscillation frequency $f$ can be
transformed into the carrier density by $n = efD/h$, where $D$ represents the
degeneracy of states. Assuming a double-degenerate
state ($D=2$), we obtained $n_{SdH}=0.98\times10^{11}\text{ cm}^{-2}$. As seen in
Fig.\,\ref{fig_par}(e), $\frac{n_{SdH}}{2}$ overlaps with the densities of
carriers $\mathbf{e_1}$ and $\mathbf{e_2}$.
As discussed below, these carriers can be attributed to bottom and top
surface states, respectively.
Note that SdH oscillations are
mostly sensitive to the carriers density $n$. In the present case,
the magnetotransport signal does not show any clear indication of the
presence of two carrier types with different densities. In fact,
Landau filling factors $v = n_{SdH}/(B_{min}e/h)$ at the minima of $\rho_{xx}$
seems to give odd values ($v=7,9,11,13,15$), which is a characteristic
signature of a double-degenerate Dirac system~\cite{neto_rmp_2009, buttner_nphys_2011, brune_prx_2014}.

To compare theory and experiment without using isotropic approximation, the
cyclotron mass can be plotted directly as a function of the 2D density.
This presentation is given in Fig.\,\ref{fig_mass} where the
$\mathbf{k \cdot p}$ predictions are shown with empty symbols and the
experimental results with full symbols.

\textcolor{black}{
The theoretical points were obtained for a discrete number of $n_{tot}$ as
discussed in Sec.~\ref{secmod}.
A scattering in the theoretical data comes from several effects:
(i) numerical integration of area $A$ in Eq.\,(\ref{eqCR}) with a discrete number of $k$points,
(ii) from the anticrossings
of the subbands, and (iii) from a finite value of the lateral lattice constant ($a=1$\,nm)
in the full-band envelope-function approach used for the self-consistent calculations.
}

We note that the
approximate density independence of the majority of the observed carriers in
Figs.\,\ref{fig_mass}(a) and \ref{fig_mass}(b) suggests that the dispersion relations will have a
paraboliclike shape. Indeed, inserting $E=\hbar^2k^2/2m_c$ into
Eq.\,(\ref{eqCR2}) gives the momentum and density-independent cyclotron mass
$m_c=eB/\Omega_c = \mathrm{const}(n,k_F)$. However, since hybridization of
multiple subbands takes place in the system, we do not expect a simple
parabolic band structure, but one with higher-order corrections.

Comparing the experimental points [solid symbols in Fig.~\ref{fig_mass}(a)]
with theoretical predictions, we recognize the $\mathbf{h_2}$ carriers
as the fingerprint of the first spin-polarized valence band
with
four degenerate islands pockets in the band dispersion.
Apparently, within the gate-voltage range of the present experiment, we did
not reach the region of the ringlike Fermi surface nor the
rest of the valence subbands at lower energies.
Most likely, this is
due to the flatness of the band structure at the transition point which leads
to small values of $\partial E_f / \partial U_g$.
Experiments in quantizing magnetic fields previously showed a
transition line involving the
hole Landau level in a $20$-nm sample~\cite{Zholudev2012}. Nevertheless,
our results exhibit the first detection of a hole carrier in a 3D TI by
quasiclassical cyclotron resonance analysis.

Experimental values show fairly flat behavior of the hole
mass vs its density, whereas the theoretical values are slightly increasing.
The weak increase was as well experimentally and theoretically observed for much
thinner samples ($d\leq20$~nm)\cite{Minkov2017}.
Currently,
the reasons behind the mismatch between experiment and theory are not
clear. Two factors can impact the
experimental values here: (i) The experimental data
for holes were obtained at high values of gate, which leads to deformation of the band structure, and (ii) due to relatively low hole
density and high magnetic fields ($\sim 1.5$\,T)
we are approaching the limit where transitions between
single Landau levels start to dominate.

Figure \ref{fig_mass}(b) shows the comparison of the cyclotron mass of the
electronlike carriers with model calculations.
We start with the analysis of the theoretical mass-density relations of the
surface states that are marked by TS and BS.
The density of the TS states (yellow open squares) can be changed by applying
the gate voltage within the full range of Fig.\,\ref{fig_mass}(b).
We observe approximate density independence of the cyclotron mass for the
top surface states supporting the paraboliclike form of the surface band.
We interpret the experimentally determined carriers $\mathbf{e_2}$
(yellow full squares) as the top surface carriers as their parameters are
close to the results of the theory.

On the contrary, the theoretical model predicts only weak variation of the
electron density at the bottom surface as a function of doping [green open
triangles, magnified part of Fig.\,\ref{fig_mass}(b)], which is due to
screening of the potential by the top surface (see also
Fig.~\ref{fig_hartree}). In the same mass range, we
observe the carriers $\mathbf{e_1}$ (green full triangles) that probably
correspond to the electrons on the bottom surface. Looking back at
Fig.\,\ref{fig_par}(d), we observe that the carriers $\mathbf{e_3}$ are
possibly a continuation of $\mathbf{e_1}$. Therefore, we interpret
$\mathbf{e_3}$ as the bottom surface carriers as well. The gap between these
two carriers in Fig.\,\ref{fig_par}(d) might be the result of a dominating
cyclotron signal by the top surface carriers $\mathbf{e_2}$ at the gate
voltages between $+1$ and $+7$\,V.

\begin{figure*}[tbp]
\includegraphics[width=1\linewidth]{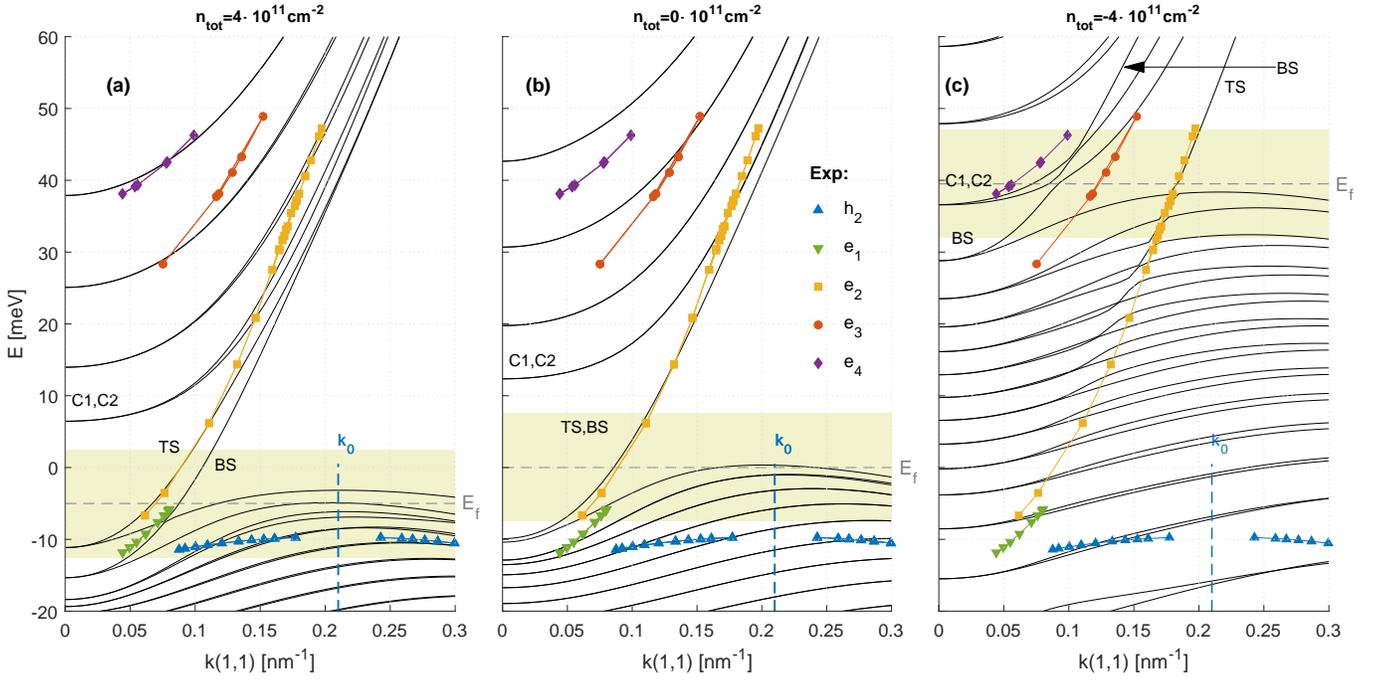}
\caption{\textit{Band structure of the three-dimensional topological insulator
based on a strained HgTe along the (1,1) direction.} Symbols: experimental data obtained from cyclotron
mass; solid lines are predictions of the $\mathbf{k} \cdot \mathbf{p}$ model
for three values of the total charge density: (a) hole doping:
$n_{tot} = 4  \cdot  10^{11}$\,cm$^{-2}$, (b) charge neutrality:
$n_{tot} = 0$ and (c) electron doping:
$n_{tot} = - 4  \cdot  10^{11}$\,cm$^{-2}$. Areas highlighted in yellow
present the regions, where it is valid to compare experimental results with
theory.}
\label{fig_disp}
\end{figure*}

The $\mathbf{k \cdot p}$ model predicts that the bottom of the bulk
conduction band can be reached at high electron densities. For such high
voltages, not only the top and bottom surfaces are strongly split, but also
the spin degeneracy of the conduction band is lifted [see
Fig.\,\ref{fig_disp}(c)]. Thus the theory predicts two cyclotron
resonances from the bulk conduction band in the relevant doping range that
are shown in Fig.\,\ref{fig_mass}(b) by open diamonds~(C1) and open
circles~(C2). Comparing these predictions with the experiment, we suggest
that carriers $\mathbf{e_4}$ correspond to bulk conduction electrons C1.


Finally, we note that within an alternative description the electron-like
signals $\mathbf{e_3}$ and $\mathbf{e_4}$ could be identified as C1 and C2
especially since they were observed simultaneously as soon as the Fermi level
in the system reached the conduction bands. However, this interpretation
provides a less convincing agreement between theory and experiment.

\subsection{Band structure of 3D TI}

To access the experimental band structure of the HgTe film, the charge
density of electrons is transferred to the electron momentum using the relation
$k=\sqrt{4\pi n/D}$ 
~\cite{shuvaev_prb_2017}. According to the identification of the
carriers in the previous section, we assume single degeneracy ($D=1$) for
all electronlike carriers.
We classified $\mathbf{h_{2}}$ holecarriers as fourfold valley degenerate
and spin-polarized hole-pocket states, thus, taking $D=4$. According to the
model calculations, the four local maxima of the valence band are expected
at finite wave-vector $k_0\approx (\pm 0.15, \pm 0.15)\,\text{~nm}^{-1}$.
The maximum of the experimental valence band has been shifted by this value.
As pointed out in Sec.~\ref{secmod}, for holelike carriers we calculate the $k$ vector along the (1,1) direction as $k=k_0 \pm k_{eff}$ with $k_{eff}=\sqrt{\pi n}$.

The band dispersion, calculated within the approximation above, is shown in
Fig.\,\ref{fig_disp} as solid symbols. Direct integration lacks in providing
the absolute energy position of the bands. Since we assume that the gate voltage
defines a constant Fermi level in the film, the bands are vertically aligned to each other
by referring to the gate voltage at which they were mutually detected.

\begin{figure}[tbp]
	\includegraphics[width=1\linewidth]{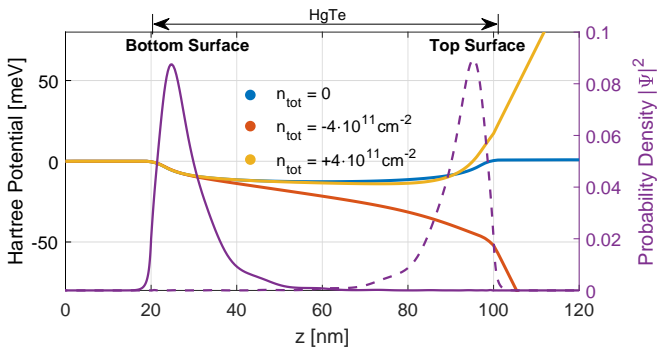}
	\caption{\textit{Hartree potential and spatial distribution of the wave functions
	for the top and bottom surface states.} Hartree potential (left axis) at
	neutral, electron and hole dopings
	that is self-consistently determined as described in Sec.\,\ref{secmod}.
	The spatial probability distribution (right axis) of the surface
	states at $n_{tot}=0$ is superimposed on the Hartree potential.
	}
	\label{fig_hartree}
\end{figure}

In Fig.\,\ref{fig_disp}, we plot the theoretical band structure for three
different doping ranges: (a) hole-doped regime with
$n_{tot}=+4\cdot10^{11}\text{~cm}^{-2}$, (b) undoped regime with $n_{tot}=0$,
and (c) electron-doped regime with $n_{tot}=-4\cdot10^{11}\text{~cm}^{-2}$.
The external electric field created by the applied gate drastically
influences the energy spectrum as seen in Fig.~\ref{fig_disp}.
This variation of the band dispersion can be well
understood taking into account the spatial distribution of the probability density of
different states and the spatial dependence of the Hartree potential.
These dependencies are shown in Fig.\,\ref{fig_hartree}. Yellow, blue, and red curves show
the Hartree potential for the same hole, neutral, and electron dopings as
in Fig.~\ref{fig_disp}(a)-\ref{fig_disp}(c). The solid and dashed violet lines represent
the probability distribution of the BS and TS states,
respectively, at $n_{tot}=0$.
While varying $n_{tot}$ does alter the distribution functions, the positions
	of the distribution maxima remain almost unchanged. Therefore,
	it is clearly evident that Hartree potential influences the top and bottom surfaces
	differently when $n_{tot}\neq0$.
It is well seen, that, at the position of
the BS, the Hartree potential barely changes with $n_{tot}$, a consequence
of the screening by all other carriers.
This explains the weak gate dependence of the BS parameters in
Fig.~\ref{fig_mass}.

On the other hand, the TS experiences the strongest
influence from the varying gate potential being easily split from the BS
and shifted in energy in the band diagram. Latter is mostly evident at the
positive gate voltages corresponding to electron doping with
$n_{tot}=-4\cdot10^{11}\text{~cm}^{-2}$. The value of the Hartree potential
at the position of the TS is around $-40$~meV. This value directly corresponds to
the shift of TS with respect to the $E_f$, when comparing the undoped and the
electron-doped regimes presented in Fig.~\ref{fig_disp}(b) and \ref{fig_disp}(c). Similar
shifting occurs for the conduction and valence bands. However, the shifting
amplitudes are smaller since the maxima of the corresponding wave functions lie
in the bulk.
The overlap between the bulk valence and TS wave functions leads to
multiple crossings and anticrossings of their dispersion curves.

The variation of the gate voltage leads to shifting of the characteristic band energies and to the
splitting of the bands that were degenerate at $n_{tot}=0$. Therefore, the
comparison between experiment and theory is valid in the vicinity of
$E_f$ only with $E_f = E_f(n_{tot})$ being the Fermi level of the system.
Three regions around Fermi energy corresponding to electron doping,
hole doping, and the charge neutrality are shown by yellow shaded areas
in Fig.\,\ref{fig_disp}(a) and \ref{fig_disp}(c), and they are defined by $E = E_f \pm 7.5$~meV.
Also, $U_g=0$\,V and the charge neutrality point do not coincide
due to impurity doping -- the experimental charge neutrality was found at
$U_g=-3$\,V.

We start the discussion of the band structure with the region close to the
charge neutrality point shown in panel~(b). Here, according to theory,
the bands are spin degenerate,
and the size quantization of the valence and conduction
bands is seen.
As the Fermi energy lies in the vicinity of zero, the active states are expected to be
the surface states and the valence-band holes.
In the experiment, however, we have detected only the TS states.
These states are marked as $\mathbf{e_2}$ throughout this
paper, and their dispersion fits well to the theoretical predictions without
additional free parameters. On the other hand, $\mathbf{e_1}$ and $\mathbf{h_2}$
appeared at lower energies as the theory predicts.

In the hole-doped region, Fig.\,\ref{fig_disp}(a), we focus only on the data
at lower energies. Here both in the experiment and in the theory, we observe a clear
splitting of the surface bands. These results are denoted as TS and BS in
the model, and they correspond to $\mathbf{e_2}$ and $\mathbf{e_1}$ carriers,
respectively.
%
The reasons behind the
vertical misalignment between the experimental and theoretical valence bands and
the resulting increase of the indirect band gap remain unclear.
Within our procedure, the energy
position of the $\mathbf{h_{2}}$ state cannot be shifted as it is fixed by
the values of the gate voltage. Here, we would like to note that the inclusion of the
BIA
terms in our theoretical model resulted in an increase in the energy at which the
valence holes appear and, therefore, an even greater mismatch with the experiment (see Fig.\,S9 of the Supplemental Material in Ref.~[\onlinecite{supp}]).

In the predominantly electronic doping regime shown in Fig.\,\ref{fig_disp}(c),
the $\mathbf{e_{2}}$ surface band nicely overlaps with the theoretical top
surface band, which is subject to hybridization and crossings/anticrossings with
several valence subbands. As discussed above, we attribute the carriers
$\mathbf{e_3}$ (solid red circles) to the bottom surface band and the carriers
$\mathbf{e_4}$ to one of the spin-polarized conduction bands (marked as C1).

Summarizing this section, in addition to two spatially separated surface bands
in Fig.\,\ref{fig_disp}, the bulk valence, and conduction bands are accessed
within the present experiment.
Although the band structure is strongly influenced by the gate voltage,
we observe reasonable coincidence between the $\mathbf{k \cdot p}$ model and
the cyclotron resonance data. We recall that, in the theoretical model, all
parameters are fixed by the known film structure and by the doping level of the
layers.

\section{Conclusions}

We investigated the cyclotron resonance of the three-dimensional topological
insulator based on a HgTe film in the subterahertz frequency range. In addition to the resonances from the top and bottom surface states, separate modes are observed that correspond to bulk
conduction and valence bands.
The quasiclassical approach is utilized to analyze the parameters of the
charge carriers, which is approved by the linearity of the cyclotron
frequency in external magnetic fields.
Within this approximation, the band structure can be extracted from the gate
dependence of the magneto-optical spectra. Considering the obvious effect
of the asymmetric gating potential on the sample, the experimental band
structure agrees reasonably well with the predictions of the
$\mathbf{k\cdot p}$ model.
Especially for the case when the Fermi level is shifted to the valence band, clear deviations between theory and experiment are observed.


\section*{Acknowledgments}

We acknowledge valuable discussions with S. Tarasenko. This work was supported
by Austrian Science Funds (Grants No. W-1243, No. P27098-N27, and No. I3456-N27),
by the Russian Foundation for Basic Research (Grant No. 17-52-14007),
and by the German Research Foundation Grants No. AS 327/5-1 and No. SFB 1143
(Project No. 247310070).

%

\end{document}